\documentclass[12pt]{iopart}

\usepackage{epsfig}
\usepackage[dvips]{color}

\definecolor{Black}{named}{Black}
\definecolor{Red}{named}{Red}
\def\hMpc{\,h^{-1}{\rm Mpc}}
\def\alt{\raise0.3ex\hbox{$\;<$\kern-0.75em\raise-1.1ex\hbox{$\sim\;$}}}
\def\agt{\raise0.3ex\hbox{$\;>$\kern-0.75em\raise-1.1ex\hbox{$\sim\;$}}}

\newcommand{\lesssim}{\; ^< \!\!\!\! _\sim \;}
\newcommand{\gtrsim}{\; ^> \!\!\!\! _\sim \;}
\newcommand{\url}{\texttt}
\newcommand{\be}{\begin{equation}}
\newcommand{\ee}{\end{equation}}
\newcommand{\bea}{\begin{eqnarray}}
\newcommand{\eea}{\end{eqnarray}}

\begin{document}

\hfill DSF-37-2005, MPP-2005-121, astro-ph/0510765

\title[The footprint of LSS on UHECR distribution]{The footprint of large
scale cosmic structure on the ultra-high energy cosmic ray
distribution}

\author{A. Cuoco$^1$, R. D' Abrusco$^1$,G. Longo$^{1,2}$, G. Miele$^1$,\\
and P. D. Serpico$^3$}

\address{$^1$ Dipartimento di Scienze Fisiche, Universit\'{a} di
Napoli ``Federico II", and INFN-Sezione di Napoli, Complesso
Universitario di Monte Sant'Angelo, Via Cintia, I-80126 Napoli,
Italy}

\address{$^2$ INAF-Sezione di Napoli, via Moiariello 16, Napoli,
Italy}

\address{$^3$ Max-Planck-Institut f\"{u}r Physik
(Werner-Heisenberg-Institut)\break F\"{o}hringer Ring 6, 80805
M\"{u}nchen, Germany}

%%%%%%%%%%%%%%%%%%%%%%%%%%%%%%%%%%%%%%%%%%%%%%%%%%%%%%%%%%%%%%%%%%%%%%%%
%%%%%%%%%%%%%%%........... Abstract .................%%%%%%%%%%%%%%%%%%%
%%%%%%%%%%%%%%%%%%%%%%%%%%%%%%%%%%%%%%%%%%%%%%%%%%%%%%%%%%%%%%%%%%%%%%%%
\begin{abstract}
Current experiments collecting high statistics in ultra-high energy cosmic
rays (UHECRs) are opening a new window on the universe. In this work
we discuss a large scale structure model for the UHECR origin which evaluates
the expected anisotropy in the UHECR arrival distribution starting from a given
astronomical catalogue of the local universe. The model takes into account the
main selection effects in the catalogue and the UHECR propagation effects.
By applying this method to the IRAS PSCz catalogue, we derive the minimum
statistics needed to significatively reject the hypothesis that UHECRs trace
the baryonic distribution in the universe, in particular providing a
forecast for the Auger experiment.
\end{abstract}

\pacs{
98.70.Sa,    %Cosmic rays (including sources, origin, acceleration, and
            %interactions)
95.80.+p,    %Astronomical catalogs, atlases, sky surveys, databases,
            %retrieval systems, archives, etc.
98.65.Dx    %Superclusters; large-scale structure of the Universe
            %(including voids, pancakes, great wall, etc.)
}

\maketitle
%%%%%%%%%%%%%%%%%%%%%%%%%%%%%%%%%%%%%%%%%%%%%%%%%%%%%%%%%%%%%%%%%%%%%%%
\section{Introduction}
%%%%%%%%%%%%%%%%%%%%%%%%%%%%%%%%%%%%%%%%%%%%%%%%%%%%%%%%%%%%%%%%%%%%%%%
Almost a century after the discovery of cosmic rays, a satisfactory
explanation of their origin is still lacking, the main difficulties
being the poor understanding of the astrophysical engines and the
loss of directional information due to the bending of their
trajectories in the galactic (GMF) and extragalactic magnetic field
(EGMF).

More in detail, given the few-$\mu$G intensity of regular and turbulent
GMF, a diffusive confinement of cosmic rays of galactic origin
is expected up to rigidity
${\cal R}\equiv p\,c/Z\,e\simeq$ few~${\times} 10^{17}\,$V, $p$ being the
cosmic ray momentum, $Z$ its charge in units of the positron one, and
$c$ the speed of light.
Still at ${\cal R}\simeq$ few~${\times} 10^{18}\,$V cosmic rays
are strongly deflected, and no directional information can be extracted.
Around ${\cal R}\sim 10^{19}\,$V the regime of relatively small
deflections in the GMF starts. The transition decades ${\cal
R}\simeq 10^{17}$--10$^{19}\,$V, though not yet useful for ``directional''
astronomy, may still show a rich phenomenology (drifts,
scintillation, lensing) which is an interesting research topic of
its own~\cite{Roulet:2003rr}.

At energies above a few~${\times} 10^{19}\,$eV, which we will refer to
as the ultra-high energy (UHE) regime, protons propagating in the
Galaxy retain most of their initial direction. Provided that EGMF is
negligible, UHE protons will therefore allow to probe into the nature and
properties of their cosmic sources. However, due to quite steep CR
power spectrum, UHECRs are extremely rare (a few particles km$^{-2}$
century$^{-1}$) and their detection calls for the prolonged use of
instruments with huge collecting areas. One further constraint
arises from an effect first pointed out by Greisen, Zatsepin and
Kuzmin~\cite{Greisen:1966jv,Zatsepin:1966jv} and since then known as
GZK effect: at energies $E\gtrsim 5{\times} 10^{19}$ eV the opacity of
the interstellar space to protons drastically increases due to the
photo-meson interaction process $p+\gamma_{\rm
CMB}\to\pi^{0(+)}+p(n)$ which takes place on cosmic microwave
background (CMB) photons. In other words, unless the sources are
located within a sphere with radius of ${\cal O}$(100) Mpc, the
proton flux at $E\gtrsim 5{\times} 10^{19}$ eV should be greatly
suppressed. However, due to the very limited statistics available in
the UHE regime
(cf. Volcano Ranch~\cite{Linsley:1963},
SUGAR~\cite{Winn:1986un}, Haverah Park~\cite{Lawrence:1991cc,Ave:2000nd},
Fly's Eye~\cite{Bird:1993yi,Bird:1994wp,Bird:1994uy}, Yakutsk~\cite{Efimov91}
AGASA~\cite{Takeda:1998ps}, HiRes~\cite{Abbasi:2002ta,Abu-Zayyad:2002sf},
and, very recently, also Auger~\cite{Sommers:2005vs}), the experimental
detection of the GZK effect has not yet been firmly established.

It has to be stressed that the theoretical tools available to probe
this extremely interesting part of the CRs spectrum are still
largely inadequate: both the modelling and the data interpretation
impose either strong assumptions based on little experimental
evidence or the extrapolation by orders of magnitudes of
available knowledge. For instance, the structure and  magnitude of
the EGMF are poorly known. Only recently, magnetic fields were
included in simulations of large scale
structures (LSS)~\cite{Dolag:2003ra,Sigl:2004yk}. Qualitatively the
simulations agree in finding that EGMFs
are mainly localized in galaxy clusters and filaments, while
voids should contain only primordial fields. However, the
conclusions of Refs.~\cite{Dolag:2003ra} and~\cite{Sigl:2004yk} are
quantitatively rather different and it is at present unclear whether
deflections in extragalactic magnetic fields will prevent astronomy
even with UHE protons or not.
Another large source of uncertainty is our ignorance on the chemical
composition of UHECRs, mainly due to the need to extrapolate for
decades in energy the models of hadronic interactions. They are an
essential input for the Monte Carlo simulations used in the analysis
and reconstruction of UHECRs showers, but the predictions of such
simulations differ appreciably already in the {\it knee} region
(around 10$^{15}$ eV), even when high quality data and deconvolution
techniques are used~\cite{Antoni:2005wq}.
Future accelerator measurements of hadronic cross
sections in higher energy ranges will ameliorate the
situation, but this will take several years at least.

From now on, therefore, we shall work under the assumptions that UHE
astronomy is possible, namely: i) proton primaries, for which
$e{\cal R}=E$; ii) EGMF negligibly small; iii) extragalactic
astrophysical sources are responsible for UHECR acceleration.
Now the question arises: might one support this scenario using the
directional information in
UHECRs? A possibility favoring these hypothesis is that relatively
few, powerful nearby sources are responsible for the UHECRs, and the
small scale clustering observed by AGASA~\cite{Takeda99} may be a
hint in this direction. However, the above quoted clustering has not
yet been confirmed by other experiments with comparable or larger
statistics~\cite{Abbasi:2004vu,Revenu:2005}, and probably a final
answer will come when the Pierre Auger Observatory~\cite{Auger} will
have collected enough data. Independently on the observation of
small-scale clustering, one could still look for large scale
anisotropies in the data, eventually correlating with some known
configuration of astrophysical source candidates. In this context,
the most natural scenario to be tested is that UHECRs correlate with
the luminous matter in the ``local" universe. This is particularly
expected for candidates like gamma ray bursts (hosted more likely in
star formation regions) or colliding galaxies, but it is also a
sufficiently generic hypothesis to deserve an interest of its own.

Aims of this work are: i) to describe a method to evaluate the
expected anisotropy in the UHECR sky starting from a given catalogue
of the local universe, taking into account the selection function,
the blind regions as well as the energy-loss effects; ii)
to assess the minimum statistics needed to significatively reject
the null hypothesis, in particular providing a forecast for the
Auger experiment.
Previous attempts to address a similar issue can be found
in~\cite{Waxman:1996hp,Evans:2001rv,Smialkowski:2002by,Singh:2003xr}.
Later in the paper we will come back to a comparison with their approaches and
results.

The catalogue we use is IRAS PSCz~\cite{saunders00a}. This has
several limitations, mainly due to its intrinsic incompleteness,
but it is good enough to illustrate the main features of the
issue, while still providing some meaningful information. This
work has to be intended as mainly methodological. An extension to
the much more detailed
2MASS~\cite{jarret2000a,jarret2004} and
SDSS~\cite{SDSS,Adelman-McCarthy:2005se} galaxy catalogues is presently
investigated.

The paper is structured as follows: the catalogue and the
related issues are discussed in Section~\ref{astrodata}. In
Section~\ref{method} we describe the technique used for our
analysis. The results are discussed in Section~\ref{results},
where we compare our findings with those obtained in previous works.
In Section~\ref{concl} we
give a brief overview on ongoing research and experimental
activities, and draw our conclusions. Throughout the paper we work
in natural units $\hbar=k_B=c=1$, though the numerical values are
quoted in the physically most suitable units.

%%%%%%%%%%%%%%%%%%%%%%%%%%%%%%%%%%%%%%%%%%%%%%%%%%%%%%%%%%%%%%%%%%%%%%%
\section{Astronomical Data}\label{astrodata}
%%%%%%%%%%%%%%%%%%%%%%%%%%%%%%%%%%%%%%%%%%%%%%%%%%%%%%%%%%%%%%%%%%%%%%%
%%%%%%%%%%%%%%%%%%%%%%%%%%%%
\subsection{The Catalogue}\label{cat}
%%%%%%%%%%%%%%%%%%%%%%%%%%%%

Two properties are required to make a galaxy catalogue suitable for
the type of analysis discussed here. First, a great sky coverage is
critical for comparing the predictions with the fraction of sky
observed by the UHECR experiments (the Auger experiment is observing
all the Southern hemisphere and part of the Northern one). Second, the
energy-loss effect in UHECR propagation requires a knowledge of the
redshifts for at least a fair subsample of the galaxies in the
catalogue. Selection effects both in fluxes and in redshifts play a
crucial role in understanding the final outcome of the simulations.

\begin{figure}[t]
\begin{center}
\psfig{figure=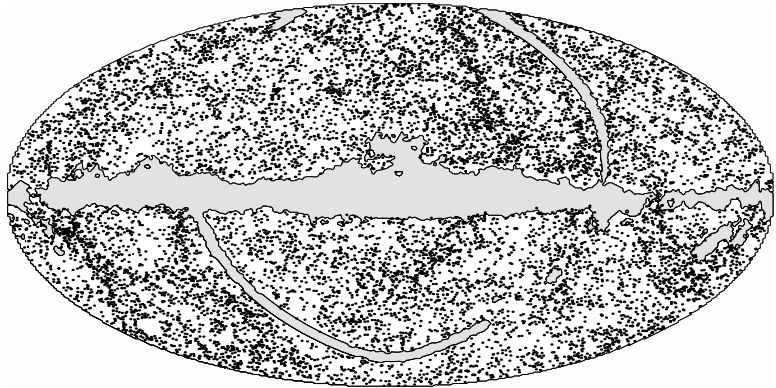,width=0.65\textwidth,angle=0}
\end{center}
\caption{\label{fig:PSCZcat} PSCz catalogue
source distribution and related mask in galactic coordinates.}
\end{figure}

Unfortunately, in practical terms this two requirements turn out to be
almost complementary and no available catalogue matches both needs
simultaneously. A fair compromise is offered by the IRAS PSCz
catalogue~\cite{saunders00a} which contains about 15 000 galaxies and
related redshifts with a well understood completeness function down to $z
\sim 0.1$ ---i.e. down to a redshift which is comparable to the attenuation
length introduced by the GZK effect--- and a sky coverage of about 84\%.
The incomplete sky coverage is mainly due to the so called zone of
avoidance centered on the Galactic Plane and caused by the galactic
extinction and to a few, narrow stripes which were not observed with enough
sensitivity by the IRAS satellite (see Fig.~\ref{fig:PSCZcat}). These
regions are excluded from our analysis with the use of the binary mask
available with the PSCz catalogue itself.

%%%%%%%%%
%
%
%%%%%%%%%%%%%

%%%%%%%%%%%%%%%%%%%%%%%%%%%%
\subsection{The Selection Function}\label{sel.func.}
%%%%%%%%%%%%%%%%%%%%%%%%%%%%

No available galaxy catalogue is complete in volume and therefore
completeness estimates derived from the selection effects in flux
are needed. More in detail, the relevant quantity to be derived is
the fraction of galaxies actually observed at the various redshifts,
a quantity also known as the \emph{redshift selection function}
$\phi(z)$~\cite{peebles80a}. A convenient way to express $\phi(z)$
is in terms of the galaxy luminosity function (i.e. the distribution
of galaxy luminosities) $\Phi(L)$ as
\begin{equation}
\phi(z) = \frac{\int_{L_{\mathrm{min}}(z)}^\infty {\rm d}\,L \ \Phi(L)}
{\int_{0}^\infty {\rm d}\,L \  \Phi(L)}.
\end{equation}
Here $L_{\mathrm{min}}(z)$ is the minimum luminosity detected by
the survey in function of redshift. By definition, for a flux-limited survey
of limiting flux $f_{\rm lim}$, $L_{\mathrm{min}}(z)$ is given in terms of
the luminosity distance $d_{L}(z)$ as
\begin{equation}
L_{\mathrm{min}}(z) = 4 \pi {d_{L}^2(z)} f_{\mathrm{lim}}.
\end{equation}
The luminosity distance depends on the cosmology assumed, though for
small redshifts ($z \lesssim 0.1$) it can be approximated by $d_{L}(z)
\simeq z/H_0$.

\begin{figure}[t]
\begin{center}
\psfig{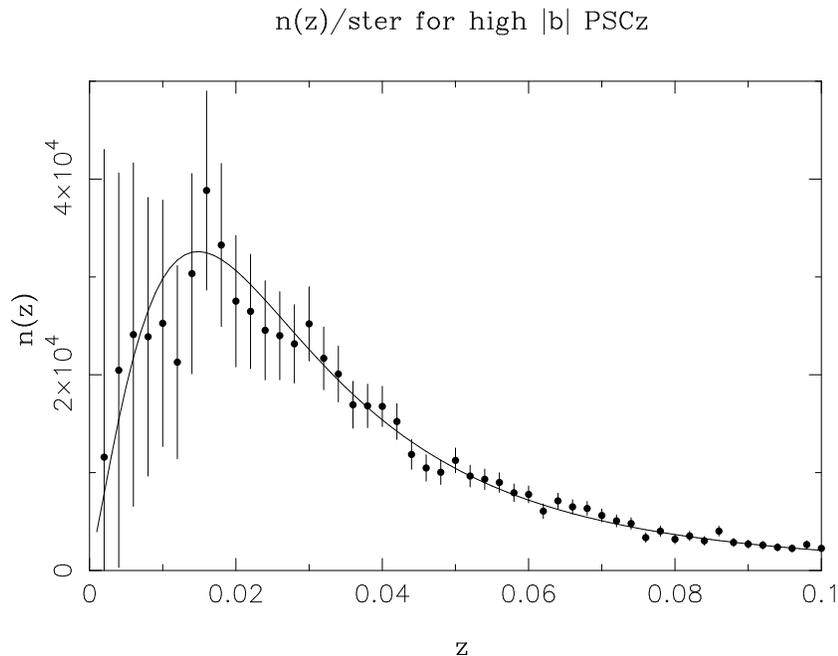}
\end{center}
\caption{\label{fig:nz} Experimental redshift
distribution of the PSCz catalogue galaxies and prediction for an
homogeneous universe from the selection function $\phi(z)$
(from~\cite{saunders00a}); both
are normalized in order to represent the number of sources per
unit of redshift per steradian.}
\end{figure}

Generally $\phi(z)$ is inferred from the catalogue data itself in a
self-consistent way, using the observational galaxy luminosity
distribution to estimate
$\Phi(L)$~\cite{saunders00a,sandage79a,efstathiou88a}.
The quantity $n(z)/\phi(z)$
represents the experimental distribution corrected for the selection
effects, which must be used in the computations. A detailed
discussion of this issue can be found in Ref.~\cite{Blanton:2000dr}.
Furthermore, we wish to stress that up to $z \sim 0.1$ evolution
effects are negligible and the local universe galaxy luminosity
function can be safely used. In the case of deeper surveys like SDSS,
cosmological effects cannot be neglected and our approach can still be
employed even though a series of corrections, like evolutionary effects
or scale-dependent luminosity, must be taken into account~\cite{tegmark03}.
These corrections are needed since luminous galaxies, which dominate
the sample at large scales, cluster more than faint
ones~\cite{davis88}.
In the case of the PSCz catalogue the selection function is given
as~\cite{saunders00a}
\begin{equation}
    \phi(r) = \phi_* {\left({\frac{r}{r_*}}\right)}^{1-\alpha}
    {\left[{1+{\left(\frac{r}{r_*}\right)}^{\gamma}}\right]}^{- \left(\beta \over \gamma \right)},
\end{equation}
with the parameters $\phi_*= 0.0077,\:\alpha =  1.82,\:r_*= 86.4,\:
\gamma=  1.56,\: \beta=  4.43$ that respectively describe the
normalization, the nearby slope, the break distance in $\hMpc$,
its sharpness and the additional slope beyond the break
(see also Fig.~\ref{fig:nz}).

%%%%%%%%%
%
%
%%%%%%%%%%%%%

It is clear, however, that even taking into account the selection
function we cannot use the catalogue up to the highest redshifts
($z \simeq 0.3$), due to the rapid loss of statistics. At high $z$,
in fact, the intrinsic statistical fluctuation due to the selection
effect starts to dominate over the true matter fluctuations,
producing artificial clusterings not corresponding to real
structures (``shot noise" effect).
This problem is generally treated constructing from the point
sources catalogue a smoothed density field $\rho(\hat{\Omega},z)$
with a variable smoothing length that effectively increases with
redshift, remaining always of size comparable to the mean distance
on the sphere of the sources of the catalogue. We minimize this effect by
being conservative in setting the maximum redshift at $z=0.06$
(corresponding to $180 \hMpc$) where we have still good statistics
while keeping the shot noise effect under control. With this threshold we
are left with $\sim 11,500$ sources of the catalogue. Furthermore, for
the purposes of present analysis, the weight of the sources
rapidly decreases with redshift due to the energy losses induced
by the GZK effect. In the energy range $E \geq\,$5${\times}
10^{19}\,$eV, the contribution from sources beyond  $z \simeq 0.06$ is
sub-dominant, thus allowing to assume for the objects beyond
$z=0.06$ an effective isotropic source contribution.

%%%%%%%%%%%%%%%%%%%%%%%%%%%%%%%%%%%%%%%%%%%%%%%%%%%%%%%%%%%%%%%%%%%%%%%
\section{The Formalism}\label{method}
%%%%%%%%%%%%%%%%%%%%%%%%%%%%%%%%%%%%%%%%%%%%%%%%%%%%%%%%%%%%%%%%%%%%%%%
In the following we describe in some detail the steps involved
in our formalism. In Sec.~\ref{propagation} we summarize our
treatment for energy losses, in Sec.~\ref{mapmaking} the way the
``effective" UHECR map is constructed, and in Sec.~\ref{statistics} the
statistical analysis we perform.

%%%%%%%%%%%%%%%%%%%%%%%%%%%%
\subsection{UHECRs Propagation}\label{propagation}
%%%%%%%%%%%%%%%%%%%%%%%%%%%%
The first goal of our analysis is to obtain the underlying
probability distribution $f_{\rm LSS}(\hat{\Omega},E)$ to have a UHECR with
energy higher than $E$ from the direction $\hat{\Omega}$.
For simplicity here and throughout the paper we shall assume that
each source of our catalogue has the same probability to emit a
UHECR, according to some spectrum at the source $g(E_i)$.
In principle, one would expect some correlation of this probability
with one or more properties of the source, like its
star formation rate, radio-emission, size,
etc. The authors of Ref.~\cite{Singh:2003xr} tested for a correlation
$L_{\rm UHECR}\propto L_{\rm FIR}^{\kappa}$, $L_{\rm UHECR}$ being the
luminosity in UHECRs and $L_{\rm FIR}$ the one in far-infrared
region probed in IRAS catalogue. The results of their analysis do
not change appreciably as long as $0\lesssim\kappa\lesssim 1$. We
can then expect that our limit of $\kappa=0$ might well work for a
broader range in parameter space, but this is not of much concern
here, since we do not stick to specific models for UHECR
sources. The method we discuss can be however easily generalized to
such a case, and eventually also to a multi-parametric modelling of
the correlation.

In an ideal world where a volume-complete catalogue were available
and no energy losses for UHECRs were present, each source should
then be simply weighted by the geometrical flux suppression $\propto
d_L^{-2}$.  The selection function already implies the change of the
weight into $\phi^{-1}d_L^{-2}$. Moreover, while propagating
to us, high-energy protons lose energy as a result of the
cosmological redshift and of the production of $e^{\pm}$ pairs and
pions (the dominant process) caused by interactions with CMB. For
simplicity, we shall work in the continuous loss
approximation~\cite{BG:1988}. Then, a proton of energy $E_i$ at the
source at $z=z_i$ will be degraded at the Earth ($z=0$) to an energy
$E_f$ given by the energy-loss equation\footnote{We are neglecting
diffuse backgrounds other than CMB and assuming straight-line
trajectories, consistently with the hypothesis of weak EGMF.}
\begin{equation}
\frac{1}{E}\frac{dE}{dz}=-\frac{dt}{dz}{\times}(\beta_{\rm
rsh}+\beta_{\pi}+\beta_{e^{\pm}}) \label{energyloss}.
\end{equation}
Eq.(\ref{energyloss}) has to be integrated from $z_i$, where the
initial Cauchy condition $E(z=z_i)=E_i$ is imposed, to $z=0$. The
different terms in Eq.~(\ref{energyloss}) are explicitly shown
below
\begin{eqnarray}\label{cosmofactor}
-\frac{dt}{dz}&=&[(1+z)H_0\sqrt{(1+z)^3\Omega_M+\Omega_\Lambda}]^{-1},\\
\beta_{\rm rsh}(z)&=&H_0\sqrt{(1+z)^3\Omega_M+\Omega_\Lambda},\\
\beta_{\pi}(z,E)&\simeq &C_\pi (1+z)^3 ,\:\:E\geq E_{\rm match} \\
{}&{}&A_\pi(1+z)^3 e^{-\frac{B_\pi}{E(1+z)}},\:\:E\leq E_{\rm
match}\\
\beta_{e^{\pm}}(z,E)&\simeq&
\frac{\alpha^3Z^2}{4\pi^2}\frac{m_e^2m_p^2}{E^3}\int_2^\infty{\rm
d}\xi \frac{\varphi(\xi)}{\exp[\frac{m_em_p\xi}{2ET_0(1+z)}]-1},
\end{eqnarray}
 where we assume for the Hubble constant $H_0=71_{-3}^{+4}$
km/s/Mpc, and $\Omega_M\simeq 0.27$ and $\Omega_\Lambda\simeq 0.73$
are the matter and cosmological constant densities in terms of the
critical one~\cite{Spergel:2003cb}. In the previous formulae, $m_e$
and $m_p$ are respectively the electron and proton masses, $T_0$ is
the CMB temperature, and $\alpha$ the fine-structure constant. Since
we are probing the relatively near universe, the results will not
depend much from the cosmological model adopted, but mainly on the
value assumed for $H_0$. More
quantitatively, the r.h.s of Eq.~(\ref{energyloss}) changes linearly
with $H_0^{-1}$ (apart for the negligible term $\beta_{\rm rsh}$),
while even an extreme change from the model
($\Omega_M=0.27$; $\Omega_\Lambda= 0.73$) to
($\Omega_M=1$; $\Omega_\Lambda= 0.0$) (the latter ruled out by
present data) would only modify the energy loss term by 6\% at
$z\simeq 0.06$, the highest redshift we consider.

The parameterization for $\beta_{\pi}$ as well as the values
$\{A_\pi,B_\pi,C_\pi\}=\{3.66 {\times} 10^{-8}{\rm yr}^{-1},2.87{\times}
10^{20}\,{\rm eV},2.42 {\times} 10^{-8}{\rm yr}^{-1}\}$ are taken
from~\cite{Anchordoqui:1996ru}, and $E_{\rm
match}(z)=6.86\,e^{-0.807\,z}{\times} 10^{20}\,$eV is used to ensure
continuity to $\beta_{\pi}(z,E)$. An useful parameterization of the
auxiliary function $\varphi(\xi)$ can be found in
\cite{Chodorowski:1992}, which we follow for the treatment of the
pair production energy loss. In practice, we have evolved cosmic
rays over a logarithmic grid in $E_i$ from $10^{19}$ to $10^{23}$
eV, and in $z$ from 0.001 to 0.3. The values at a specific source
site has been obtained by a smooth interpolation.

Note that in our calculation i) the propagation is performed to
attribute an ``energy-loss weight'' to each $z$ in order to derive a
realistic probability distribution $f_{\rm LSS}(\hat{\Omega},E)$;
ii) we are going to ``smooth" the results over regions of several
degrees in the sky (see below), thus performing a sort of weighted
average over redshifts as well. Since this smoothing effect is by
far dominant over the single source stochastic fluctuation induced
by pion production, the average effect accounted for by using a
continuous energy-loss approach is a suitable approximation.

In summary, the propagation effects provide us a ``final energy
function" $E_f(E_i,z)$ giving the energy at Earth for a
particle injected with energy $E_i$ at a redshift $z$. Note that,
being the energy-loss process obviously monotone, the inverse
function $E_i(E_f,z)$ is also available.
%%%%%%%%%%%%%%%%%%%%%%%%%%%%
\subsection{Map Making}\label{mapmaking}
%%%%%%%%%%%%%%%%%%%%%%%%%%%%
Given an arbitrary injection spectrum $g(E_i)$, the observed events
at the Earth would distribute, apart for a normalization factor,
according to the spectrum $g(E_i(E_f,z))dE_i/dE_f$. In particular we
will consider in the following a typical power-law $g(E_i)\propto
E_i^{-s}$, but this assumption may be easily generalized. Summing up
on all the sources in the catalogue one obtains the expected
differential flux map on Earth
\begin{equation}
    F(\hat{\Omega},E_f)\propto \sum_k \frac{1}{\phi(z_k)}
    \frac{\delta (\hat{\Omega}-\hat{\Omega}_k)}{4\pi d^2_L(z_k)}
    E_i^{-s}(E_f,z_k)
    \frac{dE_i}{dE_f}(E_f,z_k),
\end{equation}
where the selection function and distance flux suppression
factors have been taken into account.
However, given the low statistics of events available at this high
energies, a more useful quantity to employ is the integrated flux
above some energy threshold $E_{\rm cut}$, that can be
more easily compared with the integrated UHECR flux
above the cut $E_{\rm cut}$. Integrating the previous
expression we have
\begin{eqnarray}
    f_{\rm LSS}(\hat{\Omega},E_{\rm cut}) & \propto & \sum_k \frac{1}{\phi(z_k)}
    \frac{\delta (\hat{\Omega}-\hat{\Omega}_k)}{4\pi d^2_L(z_k)}
\int_{E_i(E_{\rm cut},z_k)}^{\infty}  \! \! \! \! \! \!
   E^{-s} {\rm d}E\nonumber  \\
       & = &  \sum_k f_{\rm LSS}(k) \ \delta (\hat{\Omega}-\hat{\Omega}_k),
\label{spikemap}
\end{eqnarray}
that can be effectively seen as if at every source $k$ of the
catalogue it is assigned a weight $f_{\rm LSS}(k)$ that takes into
account geometrical effects ($d_L^{-2}$), selection effects
($\phi^{-1}$), and physics of energy losses through the integral in d$E$.
In this ``GZK integral" the upper limit of integration is
 taken to be infinite, though the result is practically
independent from the upper cut used provided it is much larger than
10$^{20}$ eV.

\begin{figure}
\begin{center}
\begin{tabular}{c}
\psfig{figure=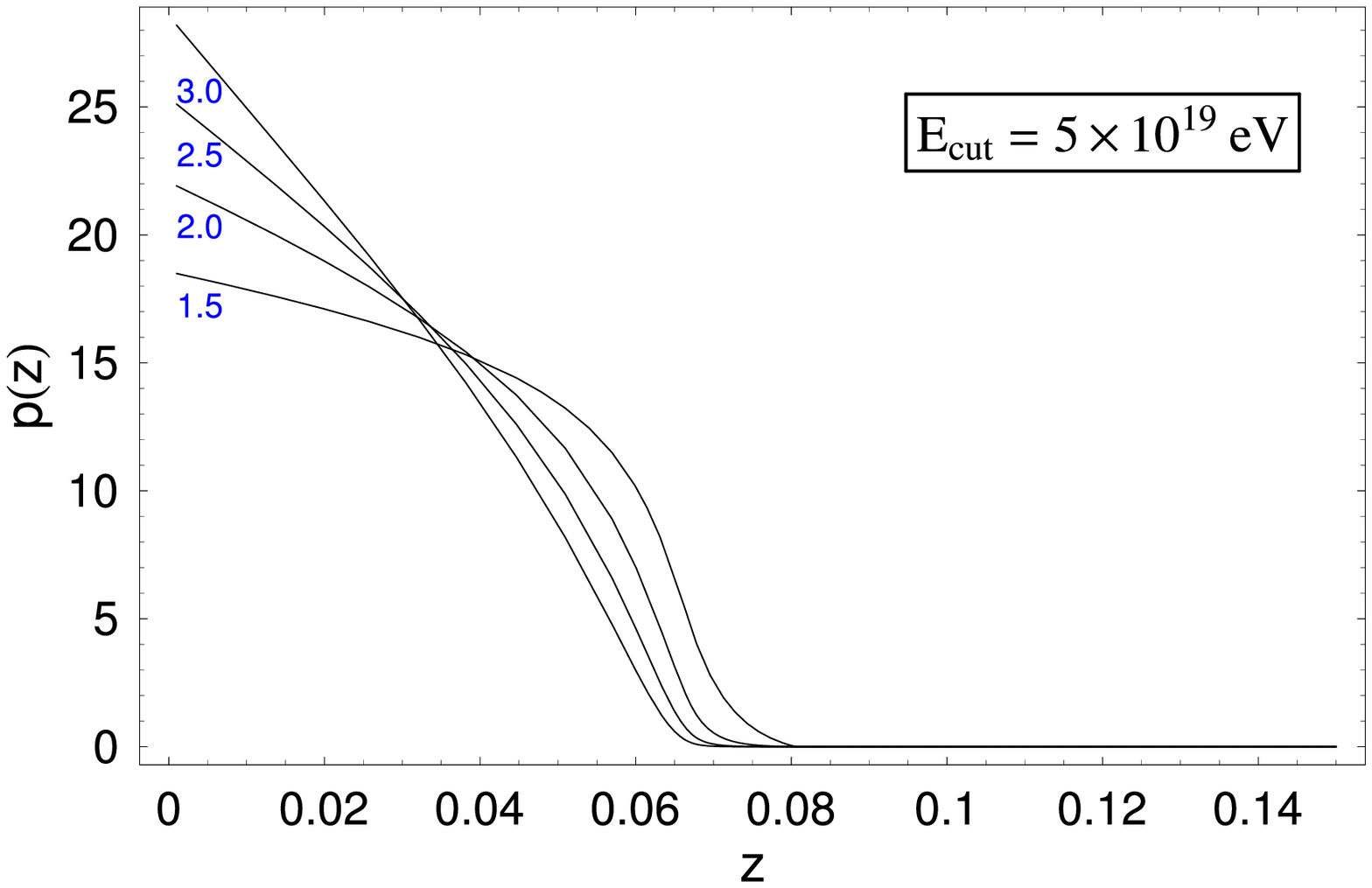,width=0.65\textwidth,angle=0} \\
\psfig{figure=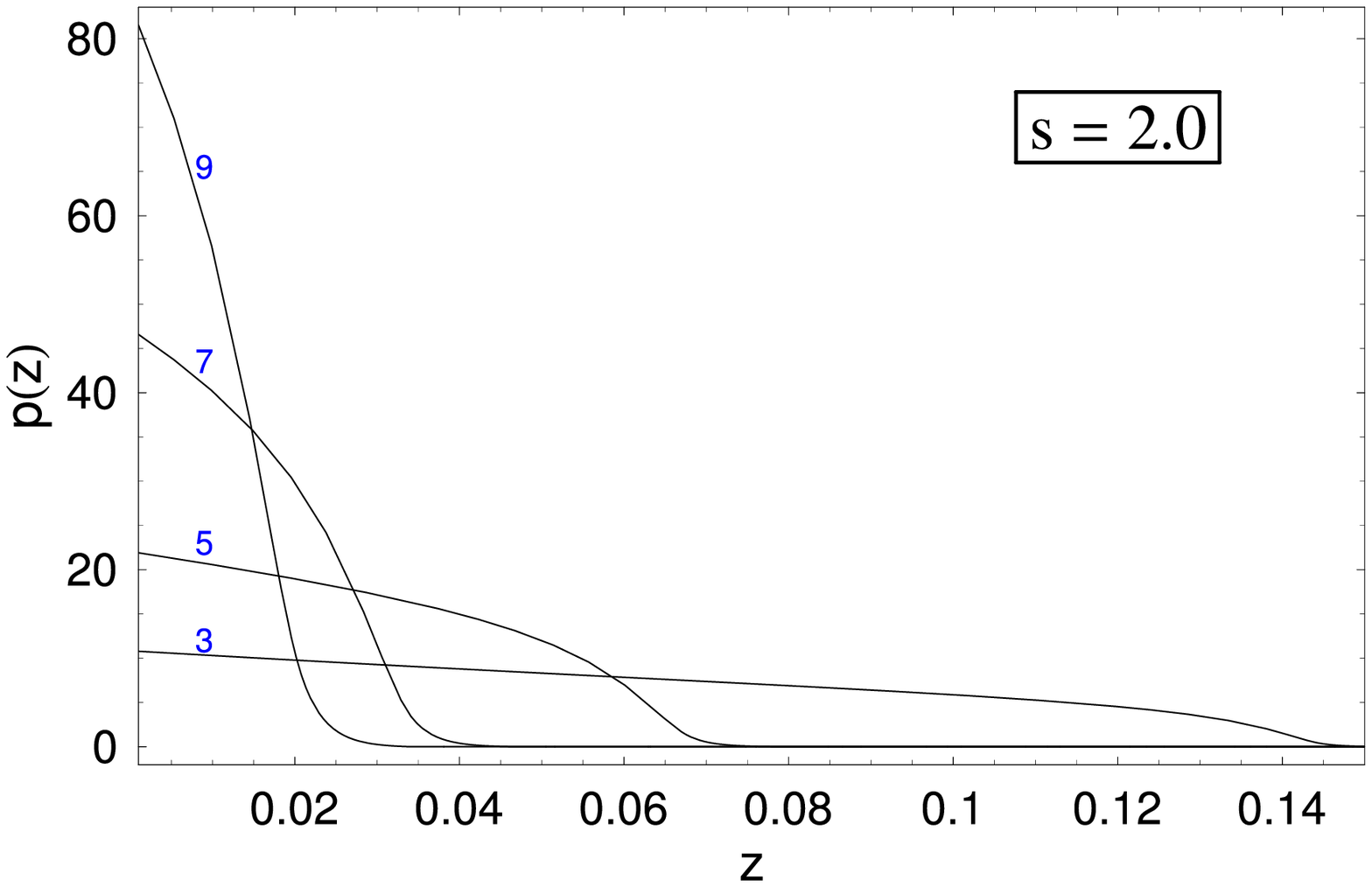,width=0.65\textwidth,angle=0}
\end{tabular}
\end{center}
\caption{\label{fig:pz_varying_slope_Ecut}
Distribution of the injection distances of CR observed at the
Earth for fixed $E_{\rm cut}=5 {\times} 10^{19}$ eV (top) and
$s=1.5,2.0,2.5,3.0$ and for fixed spectral index $s=2.0$ (bottom)
and varying $E_{\rm cut}=3,5,7,9 {\times} 10^{19}$ eV. The area subtended
by $p(z)$ has been normalized to unity.}
\end{figure}

It is interesting to compare the similar result expected for an
uniform source distribution with constant density; in
this case we have (in the limit $z\ll 1$)
\be
f_{\rm LSS}(\hat{\Omega},E_{\rm cut})\propto \int{\rm d}z\frac{\left[ E_i(E_{\rm cut},z) \right]^{-s+1}}{s-1}
\equiv\int {\rm d}z \,p(z,E_{\rm cut},s),\label{contlimit}
\ee
where the integral in d$E$ has been explicitly performed and
the flux suppression weight is cancelled
by the geometrical volume factor. The integrand $ p(z,E_{\rm cut},s)$ containing
the details of the energy losses also provides an effective cut at high $z$.
The integrand ---when normalized to have unit area--- can be interpreted
as the distribution of the injection
distances of CR observed at the Earth. It also suggests the
definition of the so-called ``GZK sphere" as the sphere from which
originates most (say 99\%) of the observed CR flux on Earth above an
energy threshold $E_{\rm cut}$. In
Fig.~\ref{fig:pz_varying_slope_Ecut} we plot the distribution $p$
for different values of $E_{\rm cut}$ and $s$. We see that around a
particular threshold $z_{\rm GZK}$ the distribution falls to zero:
the dependence of $z_{\rm GZK}$ on $E_{\rm cut}$ is quite critical
as expected, while there is also a softer dependence on $s$. This
suggests naturally the choice $E_{\rm cut}=5 {\times} 10^{19}\,$eV for
the chosen value $z_{\rm GZK} \simeq 0.06$; at the same time, the
energy cut chosen is not too restrictive, ensuring indeed that a
significant statistics might be achieved in a few years.
For this $E_{\rm cut}$ the isotropic contribution to the flux is
sub-dominant; however we can take it exactly into account and the
weight of the isotropic part is given by\footnote{The normalization
factor is fixed consistently with Eqs.~(\ref{spikemap})-(\ref{contlimit}).}
\begin{equation}
   w_{\rm iso}\propto \int_{z_{\rm GZK}}^{\infty} {\rm d}z\, p(z,E_{\rm cut})   .
\end{equation}

Finally, to represent graphically the result, the spike-like map
(\ref{spikemap}) is effectively smoothed through a gaussian filter
as
\be
\label{smoothmap}
f_{\rm LSS}(\hat{\Omega},E_{\rm cut})  \propto \sum_k f_{\rm LSS}(k) \exp
    \left(  -\frac{d_s^2[\hat{\Omega},\hat{\Omega}_k]}{2\sigma^2}
    \right)  + \frac{w_{\rm iso}}{4 \pi} 2\pi \sigma^2  \mu(\hat{\Omega})  .
\ee
In the previous equation, $\sigma$ is the width of the gaussian
filter, $d_s$ is the spherical distance between the coordinates
$\hat{\Omega}$ and $\hat{\Omega}_k$, and $\mu(\hat{\Omega})$ is the
catalogue mask (see Section~\ref{cat}) such that
$\mu(\hat{\Omega})=0$ if $\hat{\Omega}$ belongs to the mask region
and $\mu (\hat{\Omega})=1$ otherwise.

%%%%%%%%%%%%%%%%%%%%%%%%%%%%
\subsection{Statistical Analysis}\label{statistics}
%%%%%%%%%%%%%%%%%%%%%%%%%%%%
Given the extremely poor UHECR statistics,
we limit ourselves to address the basic issue of determining the
minimum number of events needed to significatively reject ``the null
hypothesis''. To this purpose, it is well known that a $\chi^2$-test
is an extremely good estimator. Notice that a $\chi^2$-test needs a
binning of the events, but differently from the K-S test performed
in \cite{Singh:2003xr} or the Smirnov-Cramer-von Mises test
of~\cite{Smialkowski:2002by}, it has no ambiguity due to the
2-dimensional nature of the problem, and indeed a similar approach
was used in~\cite{Waxman:1996hp}.  A criterion guiding in the choice
of the bin size is the following: with $N$ UHECRs events available
and $M$ bins, one would expect ${\cal O}(N/M)$ events per bin; to
allow a reliable application of the $\chi^2$-test, one has to impose
$N/M\geq 10$. Each cell should then cover at least a solid angle of
$\Delta_M\sim 10{\times}\Delta_{\rm tot}/N$, $\Delta_{\rm tot}$ being
the solid angle accessible to the experiment. For $\Delta_{\rm
tot}\sim 2\pi$ (50\% of full sky coverage), one estimates a square
window of side $454^{\circ}/\sqrt{N}$, i.e. $45^\circ$ for 100
events, $14^\circ$ for 1000 events. Since the former number is of
the order of present world statistics, and the latter is the
achievement expected by Auger in several years of operations, a
binning in windows of size $15^\circ$ represents quite a reasonable
choice for our forecast. This choice is also suggested by the
typical size of the observable structures, a point we will comment
further at the end of this Section. Notice that the GMF, that
induces at these energies typical deflections of about
$4^\circ$ \cite{KST05}, can be safely neglected for this
kind of analysis. The same remark holds for the angular resolution
of the experiment.

\begin{figure}[t]
\centering \psfig{figure=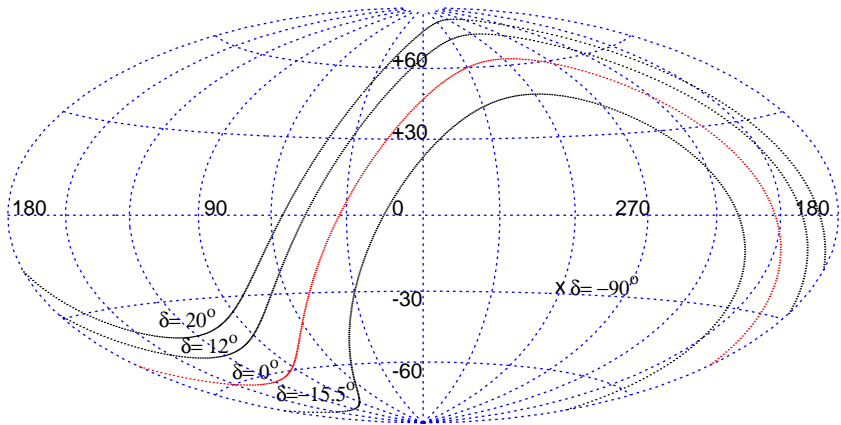,width=0.80\textwidth,angle=0}
\caption{\label{fig:RefFrame} Galactic coordinate reference frame
and contours enclosing 68\%, 95\% and 99\% of the Auger exposure
function, with the corresponding declinations. The celestial
equator ($\delta=0^\circ$) and south pole ($\delta=-90^{\circ}$) are also
shown.}
\end{figure}

Obviously, for a specific experimental set-up one must include the
proper exposure $\omega_{\rm exp}$, to convolve with the
previously found $f_{\rm LSS}$. The function $\omega_{\rm exp}$
depends on the declination $\delta$, right ascension RA, and,
in general, also on the energy. For observations having uniform
coverage in RA, like AGASA or Auger ground based
arrays, one can easily parameterize the relative exposure as
\cite{sommers2001}
\begin{equation}
\omega_{\rm exp}(\delta)\propto
\cos\theta_0\sin\alpha_m\cos\delta+\alpha_m\sin\theta_0\sin\delta,
\end{equation}
where $\theta_0$ is the latitude of the experiment
($\theta_0\approx -35^\circ$ for Auger South), $\alpha_m$ is given
by
\begin{equation}
\alpha_m = \left\{
\begin{array}{ll}
   0\,,   & {\rm if}\,\,\,\xi > 1 \\
   \pi\,,  & {\rm if} \,\,\, \xi < -1 \\
   \cos^{-1}\xi \,, \,\, & {\rm otherwise}
\end{array} \right.
\end{equation}
and
\begin{equation}
\xi\equiv \frac{\cos \theta_{\rm max} - \sin \theta_{0}\,\,\sin
\delta}{\cos \theta_{0}\,\,\,\cos\delta}\,\,,
\end{equation}
$\theta_{\rm max}$ being the maximal zenith angle cut applied (we
assume $\theta_{\rm max}=60^\circ$ for Auger). Contour plots for the
Auger exposure function in galactic coordinates are shown in
Fig.~\ref{fig:RefFrame}.

For a given experiment and catalogue, the null hypothesis we want to test is
that the events observed are sampled ---apart from a trivial geometrical
factor---
according to the distribution
$f_{\rm LSS}\,\omega_{\rm exp}\,\mu$.
Since we are performing a forecast analysis, we will consider
test realizations of $N$ events sampled according to a random
distribution on the (accessible) sphere, i.e. according to
$\omega_{\rm exp}\,\mu$, and determine the confidence level (C.L.)
with which the hypothesis is rejected as a function of $N$.
For each realization of $N$ events we calculate the two functions
\begin{eqnarray}
{\cal X}_{\rm iso}^2(N)=\frac{1}{M-1}\sum_{i=1}^{M}\frac{(o_i-\epsilon_i[f_{\rm
iso}])^2}{\epsilon_i[f_{\rm iso}]},\\
{\cal X}_{\rm LSS}^2(N)=\frac{1}{M-1}\sum_{i=1}^{M}\frac{(o_i-\epsilon_i[f_{\rm
LSS}])^2}{\epsilon_i[f_{\rm LSS}]},
\end{eqnarray}
where $o_i$ is the number of ``random" counts in the $i$-th bin $\Omega_i$,
and $\epsilon_i[f_{\rm LSS}]$ and $\epsilon_i[f_{\rm iso}]$
are the theoretically expected number of events
in $\Omega_i$ respectively for the LSS and isotropic distribution. In formulae
(see Eq.~(\ref{spikemap})),
\begin{eqnarray}
\epsilon_i[f_{\rm LSS}] & = & N  \alpha \frac{\sum_{j\in\Omega_i}f_{\rm
LSS}(j)\omega_{\rm exp}(\delta_j)\mu(j)+ w_{\rm iso}/4\pi \,
S[\Omega_i]}{\sum_{j}f_{\rm LSS}(j)\omega_{\rm exp}(\delta_j)\mu(j) +
w_{\rm iso}/4\pi \, S_{\omega}},  \\
\epsilon_i[f_{\rm iso}] & = & N  \alpha \frac{ S[\Omega_i]}{S_{\omega}},
\end{eqnarray}
where $S[\Omega_i]=\int_{\Omega_i} \! \! d\Omega \,\omega_{\rm
exp}\mu$ is the spherical surface (exposure- and mask-corrected)
subtended by the angular bin $\Omega_i$,  and similarly
$S_{\omega}=\int_{4\pi} \! \! d\Omega \,\omega_{exp}\mu$. The mock
data set is then sampled ${\cal N}$ times in order to establish
empirically the distributions of ${\cal X}_{\rm LSS}^2$ and ${\cal
X}_{\rm iso}^2$,  and the resulting distribution is studied as
function of $N$ (plus eventually $s,E_{\rm cut}$, etc.).
The parameter
\begin{equation}
    \alpha \equiv \frac{\int d\Omega \ \omega_{exp}(\delta) \mu(\Omega)}{\int d\Omega \ \omega_{exp}(\delta)}
\label{alphafactor}
\end{equation}
is a mask-correction factor that takes into account the
number of points belonging to the mask region and excluded from
the counts $o_i$. Note that the random distribution is generated with
$N$ events in all the sky view of the experiment, but, effectively,
only the region outside the mask is included in the statistical analysis
leaving us with effective $N\alpha$ events to study. This is a limiting factor
due to quality of the
catalogue: With a better sky coverage the statistics is improved
and  the number of events required to asses the model can be
reduced.

As our last point, we return to the problem of choice of the bin
size. To assess its importance we studied
the dependence of the results on this parameter. For a cell side
larger than about $\sim 25^\circ$ the analysis loses much of
its power, and a very high $N$ is required to distinguish the
models and obtain meaningful conclusions. This is somewhat
expected looking at the map results that we obtain, where
typical structures have dimensions of the
order $15^\circ- 20^\circ$. A greater cell size results
effectively in a too large smoothing and a consequent lost of
information. On the other hand, a cell size below $4^\circ- 6^\circ$
makes the use of a $\chi^2$ analysis not very reliable, because of
the low number of events in each bin expected for realistic exposure times.
In the quite large interval $\sim 6^\circ-20^\circ$ for the choice
of the cell size, however, the result is almost independent
of the bin size, that makes us confident on the reliability of
our conclusions.

%%%%%%%%%%%%%%%%%%%%%%%%%%%%%%%%%%%%%%%%%%%%%%%%%%%%%%%%%%%%%%%%%%%%%%%
\section{Results}\label{results}
\begin{figure}[p]
\begin{center}
\begin{tabular}{c}
\psfig{figure=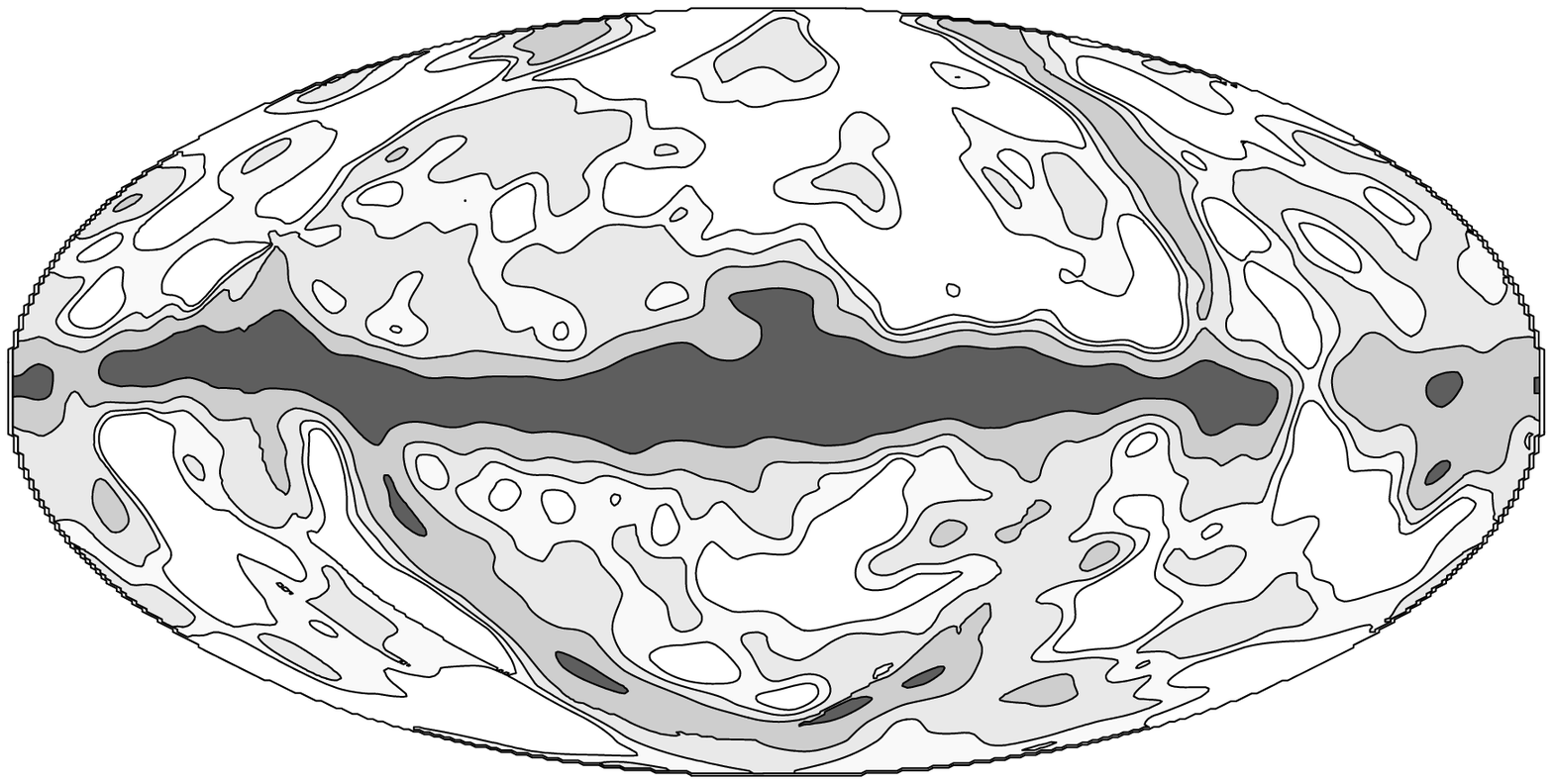,width=0.7\textwidth,angle=0} \\
\psfig{figure=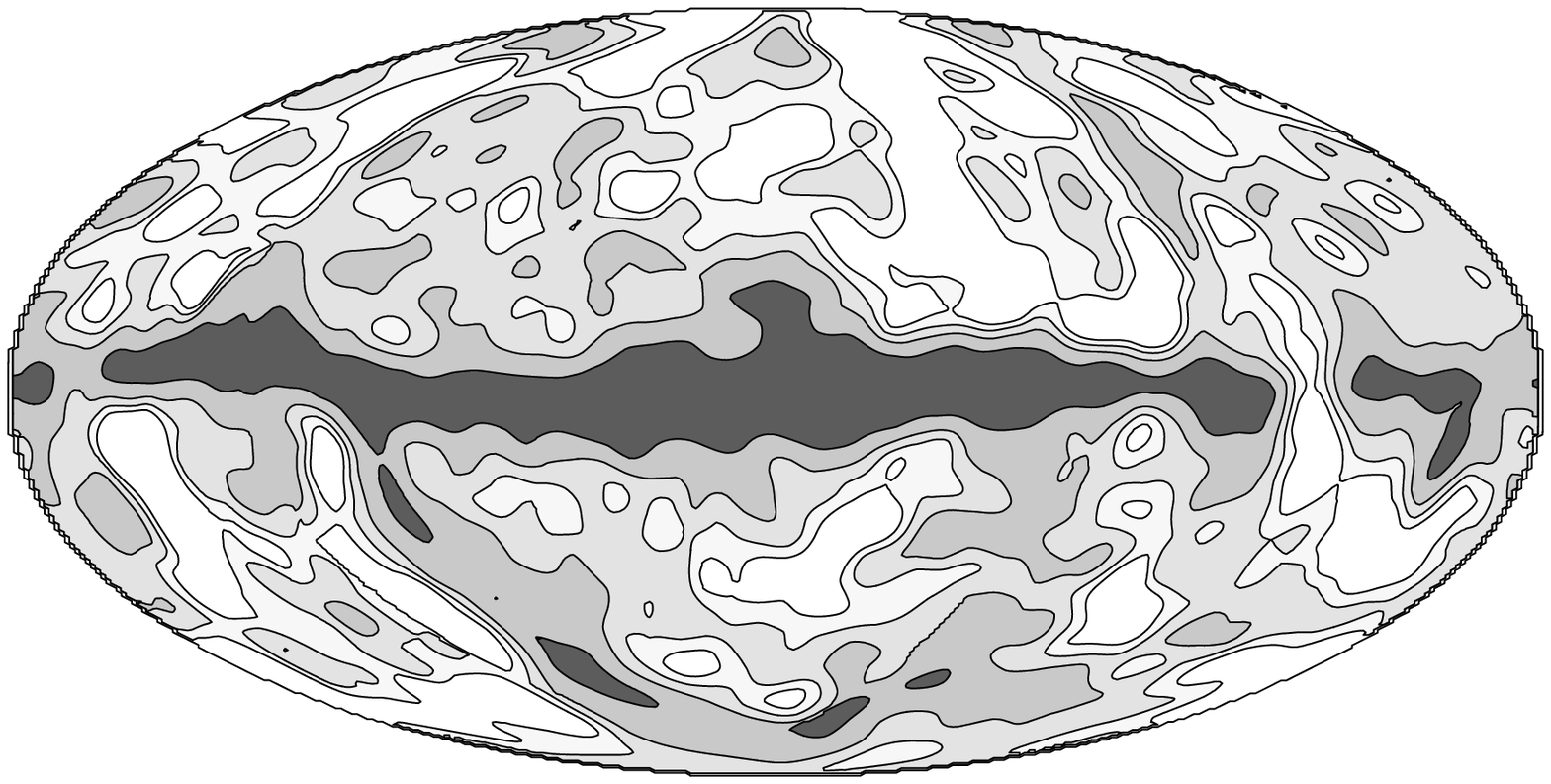,width=0.7\textwidth,angle=0} \\
\psfig{figure=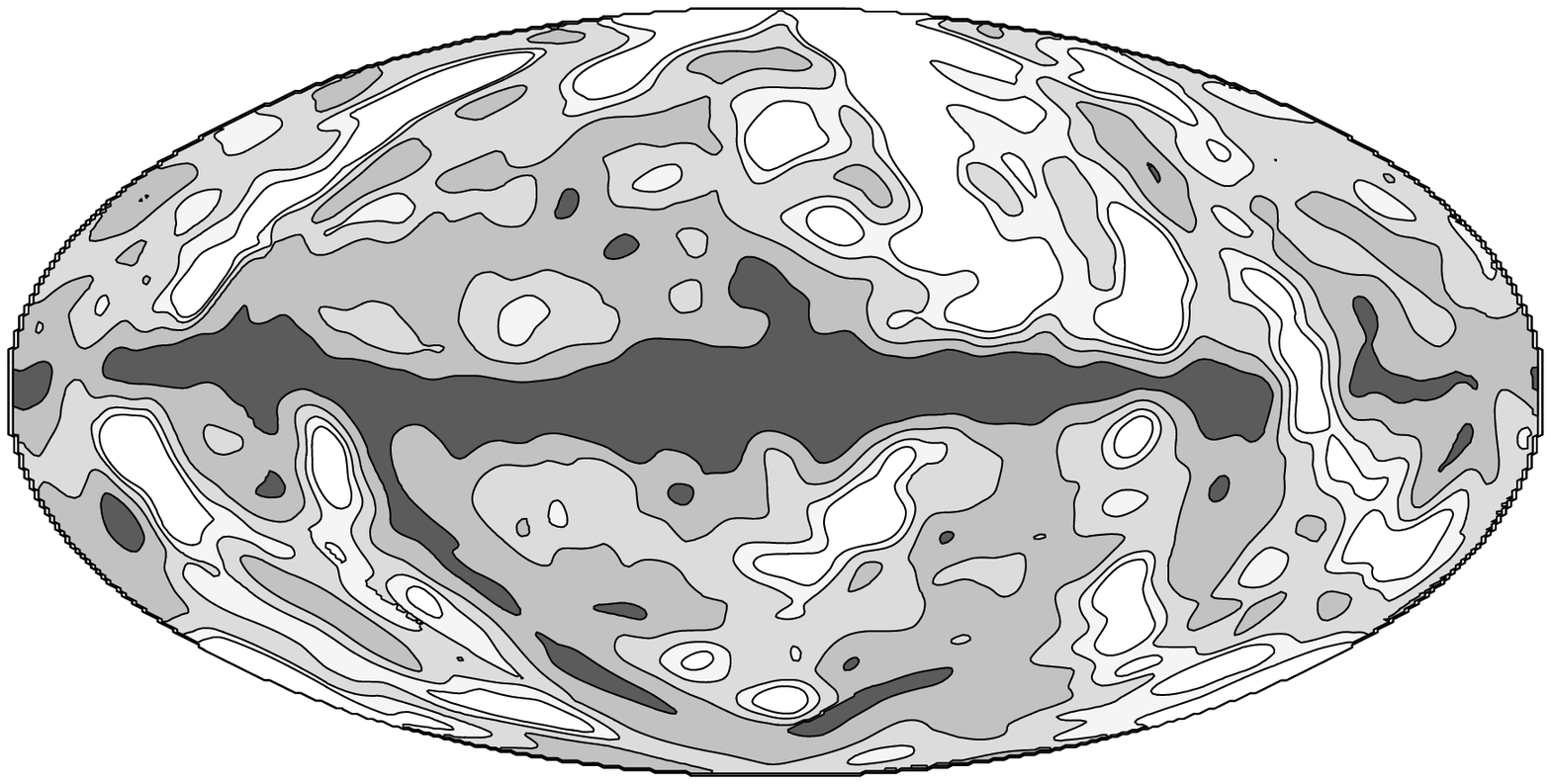,width=0.7\textwidth,angle=0} \\
\psfig{figure=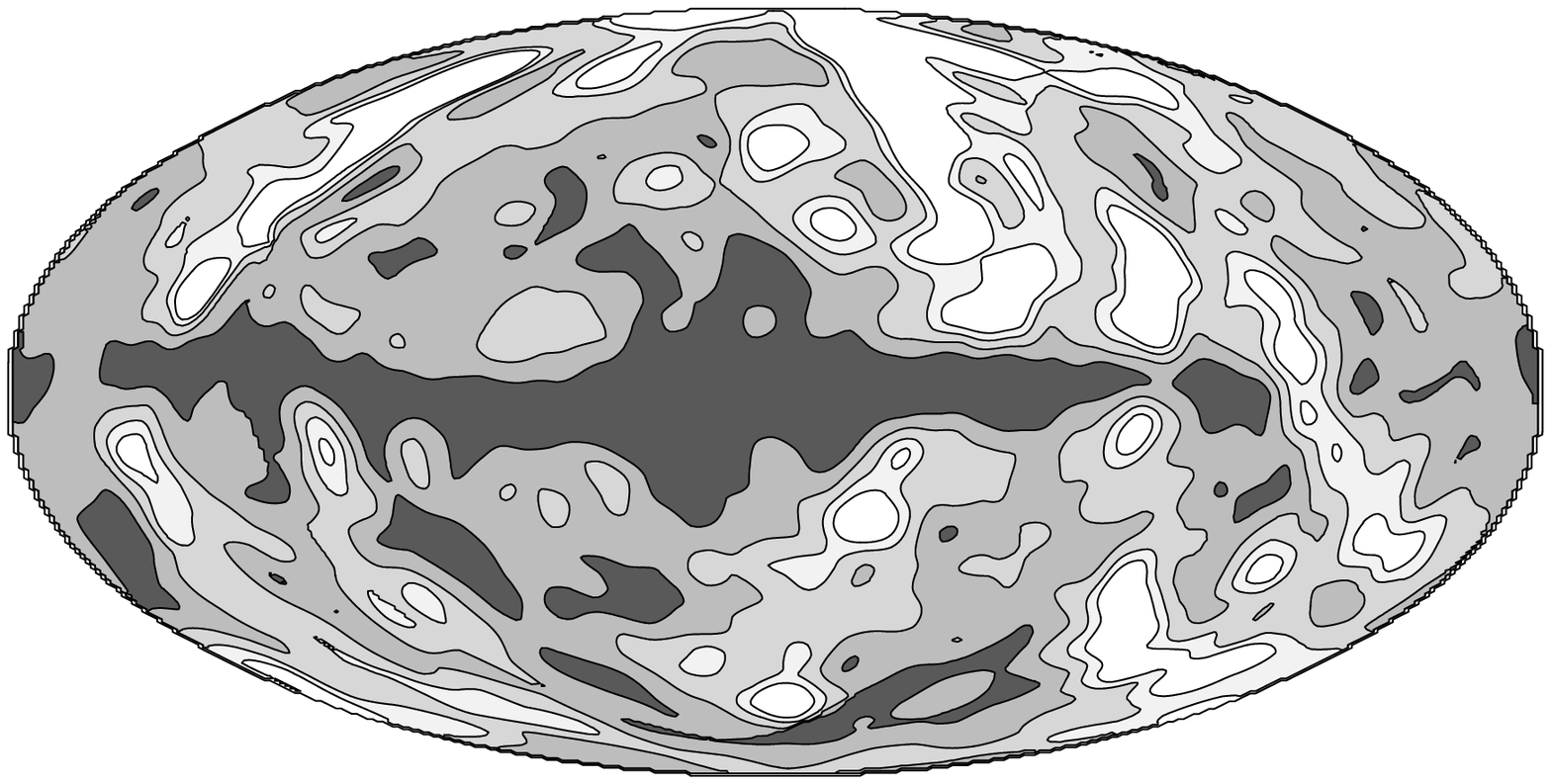,width=0.7\textwidth,angle=0} \\
\end{tabular}
\end{center}
\caption{\label{fig:ArrayMapsEcut} Equal
area Hammer-Aitoff projections of the smoothed UHECRs arrival
directions distribution (Eq.~(\ref{smoothmap})) in galactic
coordinates obtained for fixed $s=2.0$ and, from the upper to the
lower panel, for $E_{\rm cut}=3,5,7,9 {\times} 10^{19}$ eV. The
smoothing angle is $\sigma=3^{\circ}$. The contours enclose
95\%, 68\%, 38\%, 20\% of the corresponding
distribution.}
\end{figure}

%%%%%%%%%%%%%%%%%%%%%%%%%%%%%%%%%%%%%%%%%%%%%%%%%%%%%%%%%%%%%%%%%%%%%%%
In Fig.~\ref{fig:ArrayMapsEcut} we plot the smoothed maps in
galactic coordinates of the expected integrated flux of UHECRs
above the energy threshold $E_{\rm cut}=3,5,7,9{\times} 10^{19}\,$eV
and for slope parameter $s=2.0$; the isotropic part has been taken
into account and the ratio of the isotropic to anisotropic part
$w_{\rm iso}/\sum_k f_{\rm LSS}(k)$ is respectively
$83\%,3.6\%,\ll 1\%,\ll 1\%$.

Only for $E_{\rm cut}=3{\times} 10^{19}\,$eV the isotropic background
constitutes then a relevant fraction, since the GZK suppression
of far sources is not yet present.
For the case of interest $E_{\rm cut}=5{\times} 10^{19}\,$eV the contribution
of $w_{\rm iso}$ is almost negligible, while it practically disappears
for $E_{\rm cut}\agt 7{\times} 10^{19}\,$eV.
Varying the slope for $s=1.5,2.0,2.5,3.0$ while keeping $E_{\rm cut}=
5{\times} 10^{19}\,$eV fixed produces respectively the relative weights
$8.0\%,3.6\%,1.8\%,0.9\%$, so that only for very hard spectra
$w_{\rm iso}$ would play a non-negligible role
(see also Fig.~\ref{fig:pz_varying_slope_Ecut}).

\begin{figure}[t]
\centering \psfig{figure=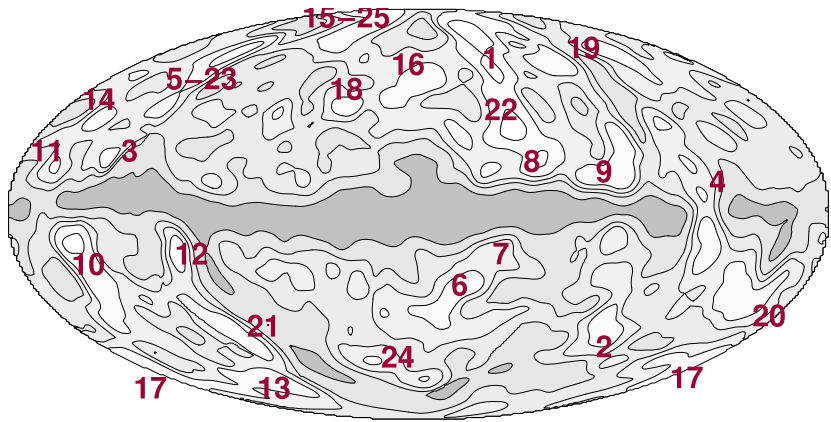,width=0.7\textwidth,angle=0}
\caption{\label{fig:legenda} Detailed key of the structures visible
in the UHECR maps; arbitrary contour levels. Labels correspond to:
(1) Southern extension of Virgo and Local Supercluster;
(2)Fornax-Eridani Cluster; (3) Cassiopea Cluster; (4) Puppis
Cluster; (5) Ursa Major Cloud; (6-7) Pavo-Indus and "Great
Attractor" region; (8) Centaurus Super-Cluster; (9) Hydra
Super-Cluster; (10) Perseus Super-Cluster; (11) Abell 569; (12)
Pegasus Cluster; (13-17) Pisces Cluster; (14) Abell 634; (15) Coma
Cluster;  (16-18) Hercules Supercluster; (19) Leo Supercluster; (20)
Columba Cluster; (21) Cetus Cluster; (22) Shapley Concentration;
(23) Ursa Major Supercluster; (24) Sculptor Supercluster; (25)
Bootes Supercluster.}
\end{figure}

Due to the GZK-effect, as it was expected, the nearest structures
are also the most prominent features in the maps. The most relevant
structure present in every slide is the Local Supercluster.
It extends along $l \simeq 140^{\circ}$ and $l \simeq 300^{\circ}$ and
includes the
Virgo cluster at $l=284^{\circ},b=+75^{\circ}$ and the Ursa Major
cloud at $l=145^{\circ},b=+65^{\circ}$, both located at $z \simeq 0.01$.
The lack of structures at latitudes from $l \simeq 0^{\circ}$ to
$l \simeq 120^{\circ}$ corresponds to the Local Void. At higher
redshifts the main contributions come from the Perseus-Pisces
supercluster ($l=160^{\circ},b=-20^{\circ}$) and the Pavo-Indus
supercluster ($l=340^{\circ},b=-40^{\circ}$), both at $z \sim
0.02$, and the very massive Shapley Concentration
($l=250^{\circ},b=+20^{\circ}$) at $z\sim 0.05$. For a more
detailed list of features in the map, see the key in
Fig.~\ref{fig:legenda}.

%%%%%%%%%%%%%
%
%
%%%%%%%%%%%%%

The $E_{\rm cut}$-dependence is clearly
evident in the maps: as expected, increasing $E_{\rm cut}$
results in a map that closely reflects
the very local universe (up to $z\sim0.03-0.04$) and its large
anisotropy; conversely, for $E_{\rm cut}\simeq3,4{\times}
10^{19}\,$eV, the resulting flux is quite isotropic and the
structures emerge as fluctuations from a background, since the
GZK suppression is not yet effective.
This can be seen also comparing the near structures
with the most distant ones in the catalogue: while the Local Supercluster
is well visible in all slides, the signal from the
Perseus-Pisces super-cluster and the Shapley concentration is of
comparable intensity only in the two top panels, while becoming
highly attenuated for $E_{\rm cut}=7{\times} 10^{19}\,$eV, and
almost vanishing for $E_{\rm cut}=9{\times} 10^{19}\,$eV.
A similar trend is observed for increasing $s$ at fixed $E_{\rm
cut}$, though the dependence is almost one order of magnitude weaker.
Looking at the contour levels in the maps we can
have a precise idea of the absolute intensity of the ``fluctuations''
induced by the LSS; in particular, for the case of interest of
$E_{\rm cut}=5{\times} 10^{19}\,$eV the structures emerge
only at the level of 20\%-30\% of the total flux, the
68\% of the flux actually enclosing almost all the sky.
For $E_{\rm cut}=7,9{\times} 10^{19}\,$eV, on
the contrary, the local structures are significantly more
pronounced, but in this case we have to face with the low
statistics available at this energies.
Then in a low-statistics regime it's not an easy task to disentangle
the LSS and the isotropic distributions.
\begin{table}
\begin{center}
\begin{tabular}{|c|cccc|}
\hline
$N\setminus s$ & 1.5 & 2.0 & 2.5 & 3.0\\
\hline
50   & (42:6) & (47:8)  & (52:10) & (52:10)\\
100  & (55:9) & (60:12) & (66:14) & (69:16)\\
200  & (72:27) & (78:33) & (84:40) & (86:43)\\
400  & (92:61) & (95:72) & (97:80) & (98:83)\\
600  & (98:85) & (99:91) & (100:96) & (100:97)\\
800  & (100:95) & (100:98) & (100:99) & (100:100)\\
1000 & (100:98) & (100:100) & (100:100) & (100:100)\\
\hline
\end{tabular}
\caption{The probability (in \%) to reject the isotropic hypothesis
at (90\%:99\%) C.L. when UHECRs follow the LSS distribution, as a
function of the injection spectral index and of the observed number
of events, fixing $E_{\rm cut}=5{\times} 10^{19}\,$eV.
  \label{table1}}
\end{center}
\end{table}

The structures which are more likely to be detected by Auger (see
also Fig.~\ref{fig:RefFrame}) are
the Shapley concentration, the Southern extension of the Virgo
cluster, the Local Supercluster and the Pavo-Indus super-cluster.
Other structures, such as the Perseus-Pisces supercluster and the
full Virgo cluster are visible only from the Northern hemisphere
and are therefore within the reach of experiments like Telescope
Array~\cite{Arai:2003dr}, or the planned North extension of the Pierre
Auger Observatory. Moreover, the sky region obscured by
the heavy extinction in the direction of the Galactic Plane reflects
a lack of information about features possibly ``hidden'' there.
Unfortunately, this region falls just in the middle of the Auger field of
view, thus reducing ---for a given statistics $N$--- the significance of
the check of the null hypothesis. Numerically, this translates into
a smaller value of the factor $\alpha$ of Eq.~(\ref{alphafactor})
with respect to an hypothetical ``twin'' Northern Auger experiment.
\begin{figure}
\begin{center}
\begin{tabular}{cc}
\psfig{figure=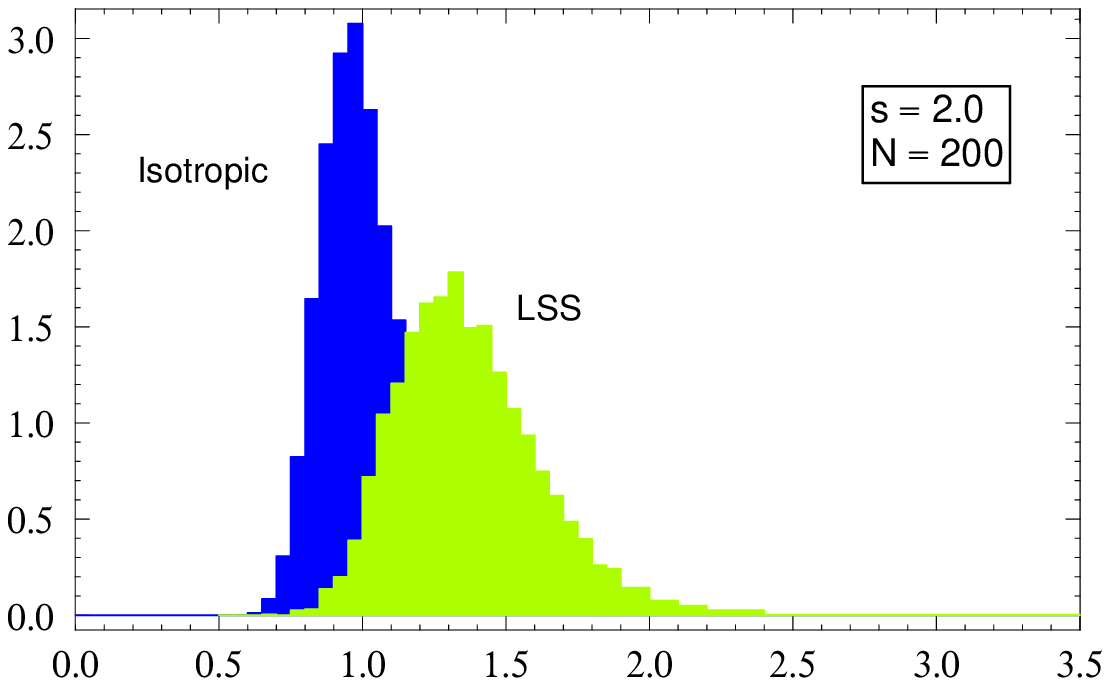,width=0.5\textwidth,angle=0} &
\psfig{figure=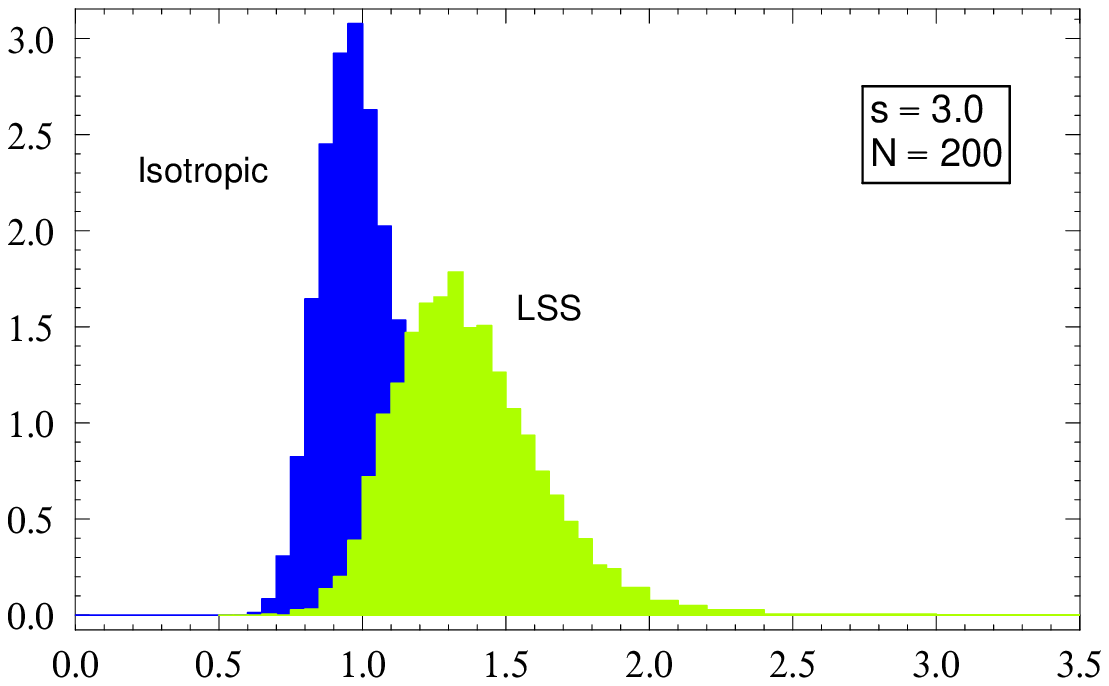,width=0.5\textwidth,angle=0}\\
\psfig{figure=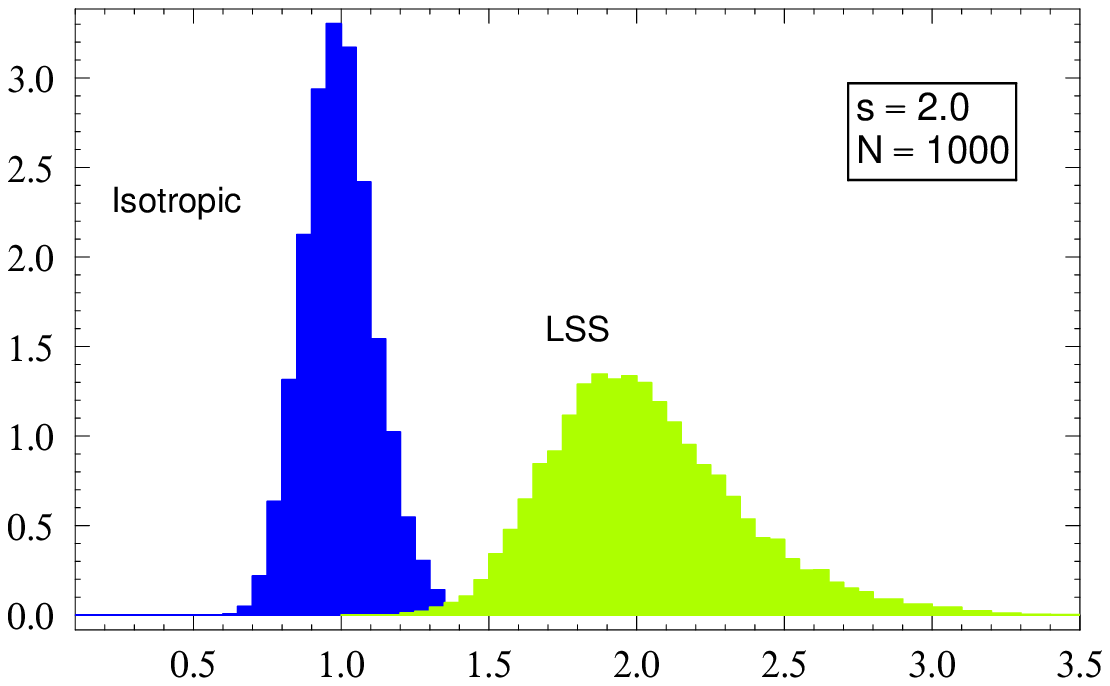,width=0.5\textwidth,angle=0} &
\psfig{figure=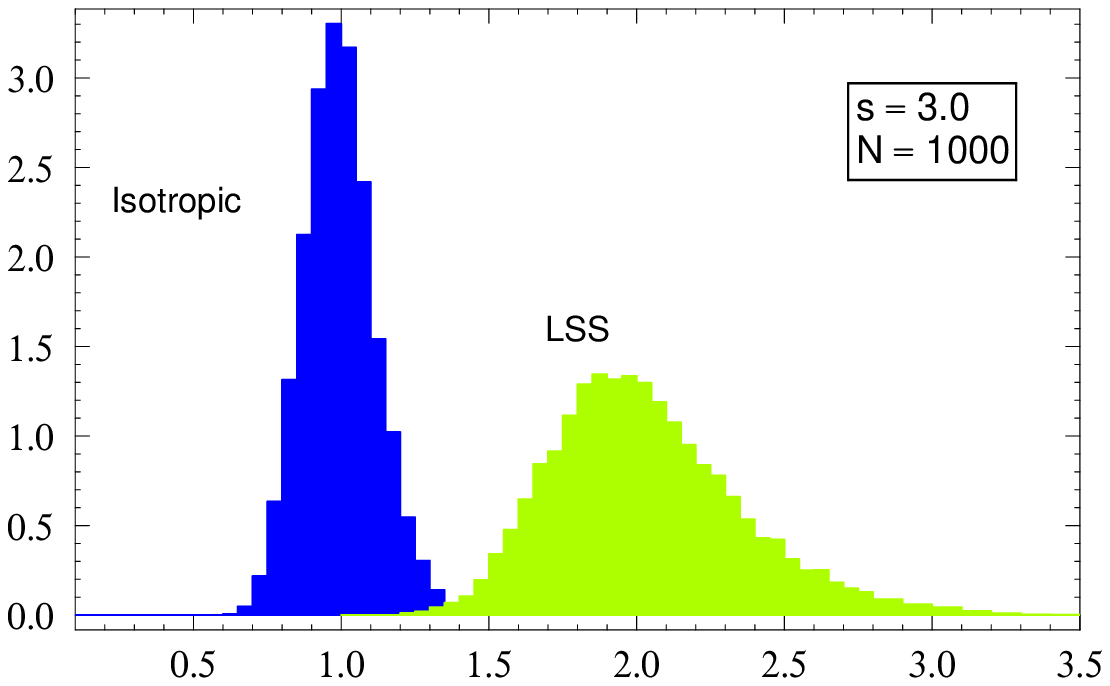,width=0.5\textwidth,angle=0}\\
\end{tabular}
\end{center}
\caption{\label{fig:chi2_nu} The probability distributions of the estimators
${\cal X}_{\rm iso}^2$ and ${\cal X}_{\rm LSS}^2$ for the cases $s=2.0,3.0$
and for $N=200,1000$ events, fixing $E_{\rm cut}=5{\times} 10^{19}\,$eV.
The distribution are the results of 10000 monte-carlo simulation like
described in the text.}
\end{figure}

A quantitative statistical analysis confirms previous qualitative
considerations. In Table~\ref{table1} we report
the probability to reject the isotropic hypothesis
at 90\% and 99\% C.L. when UHECRs follow the LSS distribution,
as a function of the injection spectral index and of the observed
number of events, fixing $E_{\rm cut}=5{\times} 10^{19}\,$eV.
In Figure~\ref{fig:chi2_nu} we show the
distributions of the functions ${\cal X}_{\rm iso}^2$ and
${\cal X}_{\rm LSS}^2$ introduced in the previous
section for $s=2.0,3.0$ and $N=200,1000$, for the same
cut $E_{\rm cut}=5{\times} 10^{19}\,$eV.
It is clear that a few hundreds events are hardly enough
to reliably distinguish the two models,
while $N=\,$800--1000 should be more than
enough to reject the hypothesis at 2-3$\,\sigma$, independently of
the injection spectrum. Steeper spectra however slightly reduce the
number of events needed for a given C.L. discrimination.
It is also interesting to note that, using different techniques
and unconstrained LSS simulations, it was found that
a comparable statistics is needed to probe a magnetized
local universe~\cite{Sigl:2004yk}.
It is worthwhile stressing that our conclusions should be
looked as conservative, since only proton primaries have been assumed,
and constant source properties. Variations in individual source
power and a mixed composition could increase the ``cosmic variance''
and make more difficult to distinguish among models for the
source distribution~\cite{Sigl:2004yk}.

With respect to previous literature on the subject, our analysis is
the closest to the one of Ref.~\cite{Waxman:1996hp}. Apart for technical
details, the greatest differences with respect to this work arise because of
the improved determination of crucial parameters undergone in the last
decade. Just to mention a few, the Hubble constant used in \cite{Waxman:1996hp}
was 100 km s$^{-1}$ Mpc$^{-1}$, against the presently determined value of
$71_{-3}^{+4}$ km s$^{-1}$ Mpc$^{-1}$: this changes by a 30\%
the value of the quantity $z_{\rm GZK}$ (see Sec.~\ref{mapmaking}).
Moreover, the
catalogue \cite{fisher95} that was used in \cite{Waxman:1996hp}
contains about 1/3 of the objects we are considering,
has looser selection criteria and larger contaminations \cite{saunders00a}.
Finally, the specific location of the Southern Auger observatory was not taken into
account. All together, when considering these factors,
we find quite good agreement with their results.

Some discrepancy arises instead with the results of
\cite{Singh:2003xr}, whose maps appear to be dominated by
statistical fluctuations, which mostly wash away physical
structures. This has probably to be ascribed to two effects, the
energy cut $E_{\rm cut}=4 {\times} 10^{19}\,$eV and the inclusion of
high redshift object (up to $z\sim 0.3$) of the catalogue
\cite{saunders00a} in their analysis. Their choice of $E_{\rm cut}=4
{\times} 10^{19}\,$eV implies indeed $z_{\rm GZK} \simeq 0.1$, i.e. a
cutoff in a redshift range where shot noise distortions are no
longer negligible. The same remarks hold for
Ref.~\cite{Smialkowski:2002by}, which also suffers of other missing
corrections~\cite{Singh:2003xr}. Also, in both cases, the
emphasis is mainly in the analysis of the already existing AGASA
data than in a forecast study. Our results however clearly show that
AGASA statistics ---only 32 data at $E\geq 5{\times} 10^{19}\,$eV in the
published data set~\cite{AGASA}, some of which falling inside the mask---
is too limited to draw any firm conclusion on the hypothesis considered.

%%%%%%%%%%%%%%%%%%%%%%%%%%%%%%%%%%%%%%%%%%%%%%%%%%%%%%%%%%%%%%%%%%%%%%%
\section{Summary and conclusion}\label{concl}
%%%%%%%%%%%%%%%%%%%%%%%%%%%%%%%%%%%%%%%%%%%%%%%%%%%%%%%%%%%%%%%%%%%%%%%
In this work we have summarized the technical steps needed to properly
evaluate the expected anisotropy in the UHECR sky starting from a given
catalogue of the local universe, taking into account the selection function,
the blind regions, and the energy-loss effects.
By applying this method to the catalogue~\cite{saunders00a}, we have
established the minimum statistics needed to significatively
reject the null hypothesis, in particular providing a forecast for the
Auger experiment. We showed with a $\chi^2$ approach that
several hundreds data are required to start testing the model at Auger
South. The most prominent structures eventually ``visible'' for this
experiment were also identified.

Differently from other statistical tools based e.g. on
auto-correlation analysis, the approach sketched above requires an
Ansatz on the source candidates. The distribution of the luminous
baryonic matter considered here can be thought as a quite generic
expectation deserving interest of its own, but it is also expected
to correlate with many sources proposed in the literature. In any
case, if many astrophysical sources are involved in UHECR
production, it is likely that they should better correlate with the
local baryonic matter distribution than with an isotropic
background.

As already stated, this work has to be intended as mainly
methodological. Until now, the lack of UHECR statistics and the
inadequacy of the astronomical catalogues has seriously limited the
usefulness of such a kind of analysis. However, progresses are
expected in both directions in forthcoming years. From the point of
view of UHECR observatories, the Southern site of Auger is almost
completed, and already taking data. Working from January 2004 to
June 2005, Auger has reached a cumulative exposure of 1750 km$^2$ sr
yr, observing 10 events over 10$^{19.7}\,$eV=5${\times}10^{19}\,$eV
(see the URL: \texttt{www.auger.org/icrc2005/spectrum.html}), Notice
that statistical and systematic errors are still quite large, and a
down-shift in the $\log_{10}E$ scale of 0.1 would for example change
the previous figure to 17 events. Once completed, the total area
covered will be of 3000 km$^2$, thus improving by one order of
magnitude present statistics in a couple of
years~\cite{Bertou:2005bx}. The idea to build a Northern Auger site
strongly depends on the possibility to perform UHECR astronomy, for
which full sky coverage is of primary importance. In any case, the
Japanese-American Telescope Array in the desert of Utah is expected
to become operational by 2007~\cite{Kasahara:2005pk}. It should offer
almost an order of magnitude larger aperture per year than AGASA in
the Northern sky, with a better control over the systematics thanks
to a hybrid technique similar to the one employed in Auger.

The other big step is expected in astronomical catalogues. The 2MASS
survey~\cite{jarret2000a} has resolved more than 1.5 million
galaxies in the near-infrared, and has been explicitly designed to
provide an accurate photometric and astrometric knowledge of the
nearby Universe. The observation in the near IR is particularly
sensitive to the stellar component, and as a consequence to the
luminous baryons. Though the redshifts of the sources have to be
obtained via photometric methods, the larger error on the distance
estimates (about 20\% from the 3-band 2MASS
photometry~\cite{jarret2004}) is more than compensated by the larger
statistics. An analysis of this catalogue for UHECR purposes is in
progress. Independently of large sky coverage, deep surveys like
SDSS~\cite{Adelman-McCarthy:2005se} undoubtedly have an important
role in mapping the local universe as well. For example, the
information encoded in such catalogues can be used to validate
methods ---like the neural
networks~\cite{Tagliaferri:2002ix,Collister:2003cz,Vanzella:2003ca}---
used to obtain photometric redshifts. An even better situation is
expected from future projects like SDSS II (see the URL:
\texttt{www.sdss.org}). Finally, a by-product of these surveys is
the discovery and characterization of active galactic nuclei~\cite{Best:2005bi,Best:2005bh}, which in turn could have
interesting applications in the search for the sources of UHECRs.

%%%%%%%%%%%%%%%%%%%%%%%%%%%%%%%%%%%%%%%%%%%%%%%%%%%%%%%%%%%%%%%%%%%%%%
%% Acknowledgments %%%%%%%%%%%%%%%%%%%%%%%%%%%%%%%%%%%%%%%%%%%%%%%%%%%
%%%%%%%%%%%%%%%%%%%%%%%%%%%%%%%%%%%%%%%%%%%%%%%%%%%%%%%%%%%%%%%%%%%%%%
\section*{Acknowledgments}
We thank M. Kachelrie{\ss} and M. Paolillo for reading the
manuscript and for useful comments. A.C. thanks the Astroparticle
and High Energy Physics Group of Valencia  for the nice hospitality
during the initial part of the work and acknowledges the
Italian-Spanish {\it Azione Integrata} for financial support.

%%%%%%%%%%%%%%%%%%%%%%%%%%%%%%%%%%%%%%%%%%%%%%%%%%%%%%%%%%%%%%%%%%%%%%%
\section*{References}
%%%%%%%%%%%%%%%%%%%%%%%%%%%%%%%%%%%%%%%%%%%%%%%%%%%%%%%%%%%%%%%%%%%%%%%

\end{document}